\documentclass[oneocolumn,preprintnumbers,amsmath,amssymb]{revtex4}

\usepackage{amsmath,latexsym,amssymb,amsfonts}
\usepackage[pdftex]{color,graphicx}
\usepackage{soul}

\begin{document}


\title{\textbf{Insights and guidelines on the Cauchy horizon theorems}}

\author{
Xiao Yan Chew$^{a}$\footnote{{\tt xychew998@gmail.com}}
and
Dong-han Yeom$^{b,c,d}$\footnote{{\tt innocent.yeom@gmail.com}}
}

\affiliation{
$^{a}$Department of Physics, School of Science, Jiangsu University of Science and Technology, Zhenjiang 212100, Jiangsu Province, People’s Republic of China \\
$^{b}$Department of Physics Education, Pusan National University, Busan 46241, Republic of Korea\\
$^{c}$Research Center for Dielectric and Advanced Matter Physics, Pusan National University, Busan 46241, Republic of Korea\\
$^{d}$Leung Center for Cosmology and Particle Astrophysics, National Taiwan University, Taipei 10617, Taiwan 
}

\begin{abstract}
Recently there has been progress to resolve the issue regarding the non-existence of the Cauchy horizon inside the static, charged, and spherically symmetric black holes. However, when we generically extend the black holes' spacetime, they are not just static but can be dynamical, thus the interior of black holes does not remain the same as the static case when we take into account the dynamical evolution of black holes. Hence, the properties of the Cauchy horizon could behave differently in the dynamical case. Then, our aim in this paper is to provide a few constructive insights and guidelines regarding this issue by revisiting a few examples of the gravitational collapse of spherically symmetric charged black holes using the double-null formalism. Our numerical results demonstrate that the inside of the outer horizon is no longer static even in late time, and the inner apparent horizon exists but is not regular. The inner apparent horizon can be distinguished clearly from the Cauchy horizon. The spherical symmetric property of black holes allows the inner horizon to be defined in two directions, i.e., the differentiation of the areal radius vanishes along either the out-going or the in-going null direction. Moreover, the Cauchy horizon can be generated from a singularity. Still, the notion of the singularity can be subtle where it can have a vanishing or non-vanishing areal radius; the corresponding curvature quantities could be finite or diverge, although the curvatures can be greater than the Planck scale. Finally, we show some examples that the ``hair" which is associated with the matter field on the inner horizon is not important to determine the existence of the Cauchy horizon; rather, the hair on the outer horizon might play an important role on the Cauchy horizon. Therefore, the dynamic properties of the interior of charged black holes could shed light for us to understand deeply about the Cauchy horizon for the extensions of no-Cauchy-horizon theorems.
\end{abstract}

\maketitle

\newpage

\tableofcontents

\section{Introduction}

The investigation of the interior of a black hole is a very interesting and important issue in general relativity. According to the singularity theorem \cite{Hawking:1970zqf}, under very reasonable and natural assumptions, the singularity must be formed as a result of a gravitational collapse. Then, one may wonder about the properties of the singularities, here we have the Reissner-Nordstrom black hole \cite{Reissner,Nordstrom} as a typical example which contains a time-like singularity and two horizons which are the outer horizon and the inner horizon. The outer horizon corresponds to the usual black hole horizon and the inner horizon corresponds to the Cauchy horizon. The Cauchy horizon preserves the predictability in general relativity where a system can be evolved with time from its given initial data which is imposed on it. When a time-like singularity exists and an observer can see an event emanating from it, then the so-called \textit{cosmic censorship conjecture} is violated \cite{Penrose:1969pc}.

Will the cosmic censorship conjecture be true in generic situations? If the electric charge of the Reissner-Nordstrom black hole is greater than its mass, its time-like singularity can become a naked singularity. However, it is not quite easy to construct such an overcharged black hole using gravitational collapses \cite{Hwang:2010im,Hansen:2013vha}. This justifies the weak cosmic censorship where there must be no observer who can see the effects from the time-like singularity; in other words, no observer can cross the Cauchy horizon which can predict the evolution of a system from the imposed initial conditions on it in the strong cosmic censorship. The violation of the strong cosmic censorship also implies the breakdown of predictability in general relativity. Will the strong cosmic censorship conjecture be true, too?

Although we do not have a definite answer yet, we have some evidence to support that the strong version of the cosmic censorship conjecture is still reasonable. In general, there should exist an inner horizon before an observer can reach the time-like singularity. Here, the inner horizon is generally unstable, because the inner horizon corresponds to an infinite blue-shift instability \cite{Simpson:1973ua}, while the outer horizon corresponds to an infinite red-shift instability. Relating these instabilities, when there is a matter fluctuation and if an observer measures the corresponding fluctuation near the inner horizon, the observed energy density will be exponentially increased. This effect is known as the \textit{mass inflation} \cite{Poisson:1990eh}. As long as mass inflation exists, we cannot trust the interior structure of the Reissner-Nordstrom spacetime anymore, and we have to rely on the full dynamical simulations \cite{Bonanno:1994ma}.

In this context, recently there are interesting papers in the literature that discuss the existence or non-existence of the Cauchy horizons \cite{An:2021plu,Cai:2020wrp,Hartnoll:2020fhc,Dias:2021afz,Devecioglu:2021xug,Grandi:2021ajl,Yang:2021civ,Henneaux:2022ijt,Luk:2022rgs,Devecioglu:2023hmn}. In the literature, some authors proved the non-existence of the Cauchy horizons by considering some matter fields which support the black holes, i.e., the existence of \textit{hairs} which refers to the extra global charges (primary hair) associated with the matter fields or only the matter fields themselves (secondary hair). However, to prove a theorem on the non-existence of the Cauchy horizons, one needs to assume some conditions to define the inner horizons or Cauchy horizons. To generalize the corresponding mathematical theorems, the underlying assumptions must be true, but one may wonder whether \textit{these assumptions will be still true in fully dynamical situations}.

The aim of this paper is not to criticize other literature but to provide useful insight to learn some interesting properties of the Cauchy horizons for the black holes in the dynamical evolution. Indeed, the Cauchy horizon and the inner apparent horizon are distinguishable in dynamic cases, and the situations become more complicated if we take into account the quantum effects. We will report the detailed results that we can learn from numerical computations. We also wish our work can shed light on the development of some advanced theorems about the Cauchy horizons of black holes in the future.

This paper is organized as follows. In Sec.~\ref{sec:pre}, we discuss some preliminary topics about the interior structure of the Reissner-Nordstrom black hole, the instability of inner horizons, and the recent progress with several literature that reformulates the no-Cauchy-horizon theorems. In Sec.~\ref{sec:les}, we mainly demonstrate the dynamical properties of the inner horizon by revisiting several models of charged black holes in the gravitational collapse using the double-null formalism which can study the full dynamical process of black holes that begins from their formation and then ends with their evaporation. Finally, in Sec.~\ref{sec:con}, we provide a few implications and guidelines that are related to the Cauchy horizon based on our numerical results and comment on possible future research directions.

\section{\label{sec:pre}Preliminaries of inner horizons}

In this section, we maximally extend the Reissner-Nordstrom black hole which is a typical example of a static and charged black hole to understand its basic interior structure, particularly the Cauchy horizon inside the outer horizon. We then discuss the instability of the inner horizon. Moreover, we discuss the recent progress that reformulates the non-existence of the Cauchy horizon by providing several references.

\begin{figure}
\begin{center}
\includegraphics[scale=0.5]{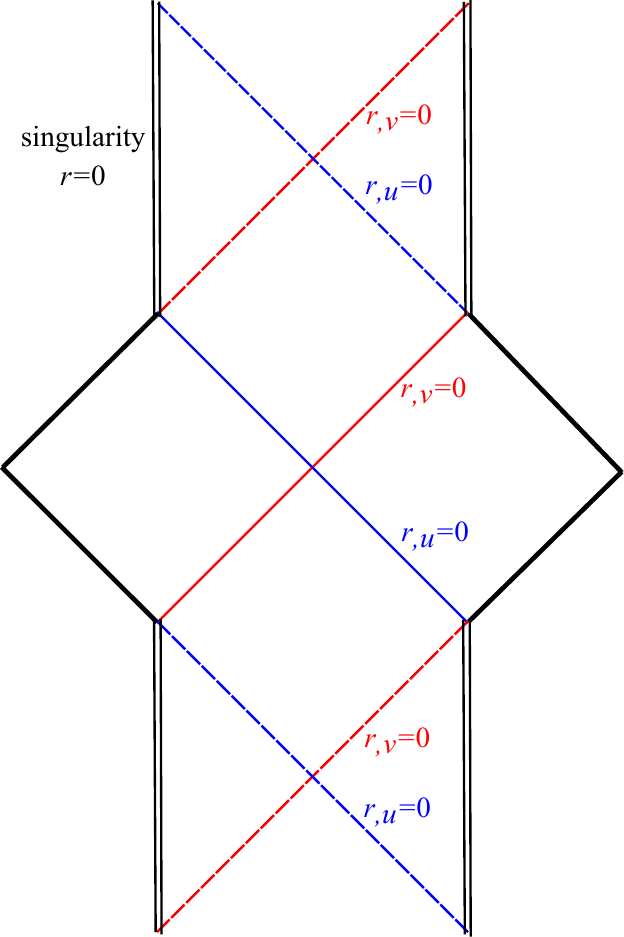}
\caption{\label{fig:charged_metric}The Penrose diagram of a static charged black hole. There exists a time-like singularity ($r = 0$). The inner and outer apparent horizons correspond to the Cauchy and event horizons. $r_{,v} = 0$ horizons (red color) are always parallel to the out-going null direction, while $r_{,u} = 0$ (blue color) horizons are always parallel to the in-going null direction.}
\end{center}
\end{figure}

\subsection{Static charged black holes}

Here we begin with the Reissner-Nordstrom black hole which is a static, stationary, and charged black hole, with the given metric as shown in the following,
\begin{eqnarray}
ds^{2} = - f(r) dt^{2} + \frac{1}{f(r)} dr^{2} + r^{2} d\Omega_{2}^{2},
\end{eqnarray}
where
\begin{eqnarray}
f(r) = 1 - \frac{2M}{r} + \frac{Q^{2}}{r^{2}},
\end{eqnarray}
where $M$ is the ADM mass and $Q$ is the asymptotically defined electric charge \cite{Reissner,Nordstrom}. Since $f(r) = 0$ can yield two solutions, one can notice that, if $M > Q$, this metric has two apparent horizons which can be denoted as $r_{+}$ and $r_{-}$ with $r_{+} > r_{-}$, where $r_{+}$ is the outer horizon and $r_{-}$ is the inner horizon.

Let us define that $u$ denotes the in-going null direction and $v$ denotes the out-going null direction. After we maximally extend the spacetime in Kruskal-Szekeres coordinates \cite{dInverno:1992gxs}, i.e.,
\begin{eqnarray}
ds^{2} = - \frac{4 r_{+}^{4} \left( r - r_{-} \right)^{1 + r_{-}^{2}/r_{+}^{2}}}{r^{2} \left(r_{+} - r_{-}\right)^{2}} e^{\frac{r_{-} - r_{+}}{r_{+}^{2}}r} dudv + r^{2} d\Omega^{2}
\end{eqnarray}
with $r = r(u,v)$, one can notice that there are two directions of the horizons, where $r_{,v} = 0$ horizons (red color) are always parallel to the out-going null direction, while $r_{,u} = 0$ horizons (blue color) are always parallel to the in-going null direction as depicted in Fig.~\ref{fig:charged_metric}. In addition, there exists a time-like singularity at $r = 0$; the inner horizon is the boundary where we cannot decide the geometry from the past data, and hence, this horizon is known as the Cauchy horizon \cite{Hawking:1973uf}. The outer horizon is the boundary between the interior and the asymptotic infinity, and hence, this horizon is known as the event horizon. So, in the static and charged black hole case, the inner apparent horizon is the Cauchy horizon and the outer apparent horizon is the event horizon.

Here, we list down several \textit{common senses} of the charged black holes:
\begin{itemize}
\item[--] $r_{,v} = 0$ apparent horizon is parallel to the out-going null direction; $r_{,u} = 0$ apparent horizon is parallel to the in-going null direction.
\item[--] The $r = 0$ singularity is time-like.
\item[--] The inner apparent horizon is the Cauchy horizon; the outer apparent horizon is the event horizon.
\end{itemize}
However, these common senses are valid only if the metric is static. In the dynamic cases, all three assertions are not valid in general. On the other hand, if one applies such common sense to some proofs about the non-existence of inner horizons, the corresponding results might not be rigorous or even wrong because, in a realistic situation, the metric should be dynamic.

\subsection{Instability of inner horizons}

One of the most important issues about the inner horizon is the instability of inner horizons. When there exists an energy flux along the in-going null direction, we can define an observer who is moving along the out-going null direction and measures the energy flux. Near the inner horizon, the observed energy flux is approximately
\begin{eqnarray}
\rho \propto e^{+\kappa v},
\end{eqnarray}
where $\kappa$ is the surface gravity of the inner horizon \cite{Poisson:1990eh}. Therefore, the observed energy must be exponentially amplified near the inner horizon. This effect is known as the mass inflation.

Theoretically, we can have a static, stationary, and charged black hole solution that is independent of time. However, in reality, there is no such an eternal object in the universe; every black hole must be generated from gravitational collapses. Therefore, energy fluctuations necessarily exist inside a charged black hole. Thus, the consequence of mass inflation is that, in realistic black holes that are generated from gravitational collapses, there must be a strong back-reaction due to the exponentially increased energy flux \cite{Bonanno:1994ma}.

Hence, it is unlikely to precisely study the interior structures of a charged black hole from the perturbative analysis on the background of static solution. The only possible way to precisely understand the interior structure is to numerically evolve the corresponding equations of motion. In this paper, we apply the double-null formalism to study the formation of an inner horizon in the gravitational collapse of charged black holes.

\subsection{No-Cauchy-horizon theorem}

In this context, there have been some mathematical investigations and proofs regarding the inner horizon of black holes. In the literature, these investigations are mainly to formulate the \textit{no-Cauchy-horizon theorems}. For example, we specify several mathematical assertions from the literature as follows: 
\begin{itemize}
\item[--] Ref.~\cite{An:2021plu} has shown numerically the Cauchy horizon does not exist in the interior of the scalar-charged hairy black holes with spherical and planar symmetries in the Born-Infield theory.
\item[--] Ref.~\cite{Cai:2020wrp} has shown numerically the Cauchy horizon does not exist in the interior of the scalar charged hairy black holes with spherical and planar symmetries in Einstein-Maxwell-scalar theory of a complex scalar field with an arbitrary form of the scalar potential.
\item[--] Ref.~\cite{Devecioglu:2021xug} has studied the non-existence of the Cauchy horizon for some specific types of black holes analytically in the Einstein-Maxwell-Horndeski theories.
\item[--] Ref.~\cite{Grandi:2021ajl} has studied numerically the non-existence of Cauchy horizon for scalar charged hairy black holes in the 5-dimensions of Einstein-Gauss-Bonnet theory with a massive complex scalar field.
\item[--] Ref.~\cite{Yang:2021civ} has studied the possibility of putting a constraint on the number of horizons for the static black holes which can only possess at most one non-degenerated inner Killing horizon inside the event horizon when they satisfy the strong or null energy condition. 
\item[--] Ref.~\cite{Devecioglu:2023hmn} has studied numerically the non-existence of Cauchy horizon for charged black holes with planar configuration in the Einstein-Maxwell-Gauss-Bonnet-scalar theory.
\end{itemize}

Here we list down some common features in the literature as follows.
\begin{itemize}
\item[--] 1. Some authors assumed that the black hole geometry will follow a \textit{regular and static solution}. If a gravitational collapse occurs and then after a long time has sufficiently passed, a black hole solution will approach a static solution, but this is in the perspective of the \textit{outside} observer. One can ask whether this assumption is still true for an inside observer.
\item[--] 2. Some authors assumed hairs to prove the theorems for some specific cases, i.e., the dynamics of matter fields near the inner horizon. If any inconsistency can arise from the corresponding dynamics, then they can prove the non-existence of the Cauchy horizon. However, the no-hair theorem has not been established near the inner apparent horizon, and thus there is no systematic way to evade the no-hair theorem inside the outer horizon; hence, it would be risky to assume the existence of such a conserved quantity associated with the symmetry of the theory for the inner horizon.
\end{itemize}
In this paper, we will show some examples that satisfy the following conditions. (1) The inner horizon is formed and approaches a (weak) curvature singularity. Hence, the interior geometry is no longer static or regular. (2) Near the inner horizon, there can be non-trivial field dynamics. This is not surprising because the interior geometry is no longer static. In the next section, we will discuss more issues that go beyond common sense.

\section{\label{sec:les}Lessons from numerical investigations}

In this section, we revisit several models of charged black holes in the gravitational collapse using the double-null simulation by demonstrating the properties of the inner horizon in the gravitational collapse. For the sake of readers, we briefly introduce the double-null formalism as the first step. Then we also briefly define the properties on the inner horizon and singularity in the double-null coordinate since they can behave differently in the gravitational collapse. Lastly, we will present and discuss the properties of the inner horizon in the gravitational collapse.

\subsection{Brief summary of the double-null formalism}

Let us consider the most generic double-null coordinates with spherical symmetry:
\begin{eqnarray}
ds^{2} = - \alpha^{2}(u,v) du dv + r^{2}(u, v) d\Omega_{2}^{2}.
\end{eqnarray}
In the \textit{the double-null formalism}, one can obtain the equations of motion and solve them numerically based on this metric with appropriate boundary conditions (see Appendix C) \cite{Nakonieczna:2018tih}. So far, plenty of numerical simulations have been carried out to investigate the dynamics of the formation of inner horizons during the gravitational collapse of the charged black holes using the double-null formalism \cite{Hod:1998gy,Hwang:2011mn,Sorkin:2001hf,Hong:2008mw,Sorkin:2001hf,Hansen:2014rua,Hansen:2015dxa,Nakonieczna:2015umf,Nakonieczna:2016iof,Chen:2015nma}. In this section, we summarize some important results that are relevant to the issue of the inner horizon by revisiting some of our previous works.
\begin{itemize}
\item[--] The authors investigated the properties of the Cauchy horizon inside the formation of a charged black hole during a gravitational collapse caused by an input pulse consisting of a complex scalar field in the Einstein-Maxwell theory \cite{Hod:1998gy,Hwang:2011mn}. The numerical results show that the inner horizon is formed after the gravitational collapse. If we turn off Hawking radiation, the inner horizon becomes a Cauchy horizon.
\item[--] The authors extended the above investigation to study the semi-classical description for the emission of Hawking radiation from a charged black hole \cite{Sorkin:2001hf,Hong:2008mw}. The inclusion of Hawking radiation allows that the inner horizon and the Cauchy horizon can be distinguished.
\item[--] The authors generalized the above two investigations from the Einstein gravity to the Brans-Dicke theory that can cover a diverse range of theories, including string-inspired models \cite{Hansen:2014rua,Hansen:2015dxa,Nakonieczna:2015umf,Nakonieczna:2016iof}, $f(R)$ gravity \cite{Hwang:2011kg,Chen:2015nma}, etc. These results show that the inner and outer horizons can be varied dynamically due to the Brans-Dicke field. Also, we can see the existence or non-existence of inner horizons relating couplings between the Maxwell field and the Brans-Dicke field.
\end{itemize}

Before we proceed to present the above results, we need to define clearly the inner horizon and singularity in the next subsection, since their appearance is highly sophisticated in the double-null simulation.

\subsection{Subtleties to define inner horizon and singularity}

The Reissner-Nordstrom black hole in the double-null coordinate contains two kinds of inner horizons: $r_{,v} = 0$ and $r_{,u} = 0$, where the former is parallel to a constant $u$ line and another is parallel to a constant $v$ line. The two horizons appear with the same areal radius, and they are well coincide with the causally defined Cauchy horizon.

However, we need to distinguish three kinds of horizons for the dynamical cases:
\begin{itemize}
\item[--] H1. $r_{,v} = 0$ inner apparent horizon.
\item[--] H2. $r_{,u} = 0$ inner apparent horizon.
\item[--] H3. Cauchy horizon (boundary where one can causally determine from the past data).
\end{itemize}

To define the third type of horizon (H3) precisely, one needs to take into account the subtleties to define the singularity and its properties. Thus, let us grasp several basic concepts about singularity as follows.
\begin{itemize}
\item[--] Type 1: it lies at $r = 0$ and the corresponding curvature quantities diverge; this kind of singularity is in the classical sense.
\item[--] Type 2: it lies in $r > 0$ and the corresponding curvature quantities diverge; this kind of singularity is also in the classical sense which is so-called a weak singularity \cite{Ori:1991zz}.
\item[--] Type 3: the radius of singularity is non-vanishing but at the Planck scale $r \simeq \ell_{\mathrm{P}}$ and the corresponding curvature quantities are also the Planck scale.
\item[--] Type 4: the radius of singularity is greater than the Planck scale $r > \ell_{\mathrm{P}}$ but the curvature quantities are at the Planck scale. This kind of singularity is known as the quantum version of the weak singularity.
\end{itemize}
Therefore, we can classify that Type 1 and 2 are classical singularities, while Type 3 and 4 are defined when there exists the Planck constant $\hbar$ for the quantum regime. Type 1, 2, and 3 will dynamically generate the Cauchy horizon because one cannot extend computations inside the singularity; however, the property of Type 4 is less known, because one may extend general relativistic computations inside the singularity. The only problem is that such an extension is inconsistent in terms of quantum gravity except for some extreme cases (e.g., introducing the large $N$-rescaling \cite{Yeom:2009zp,Kim:2013fv}).


\begin{figure}
\begin{center}
\includegraphics[scale=0.35]{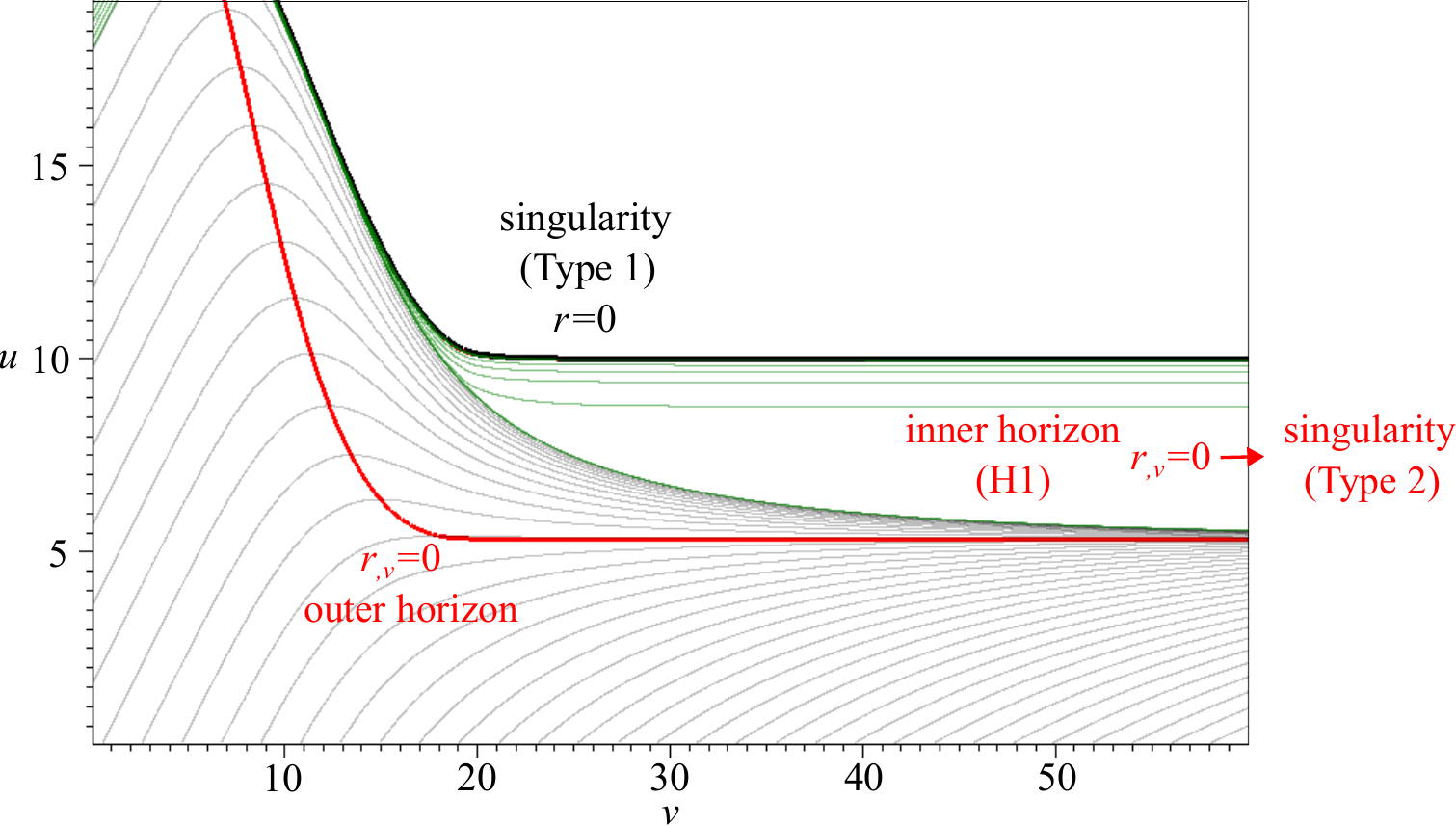}
\includegraphics[scale=0.55]{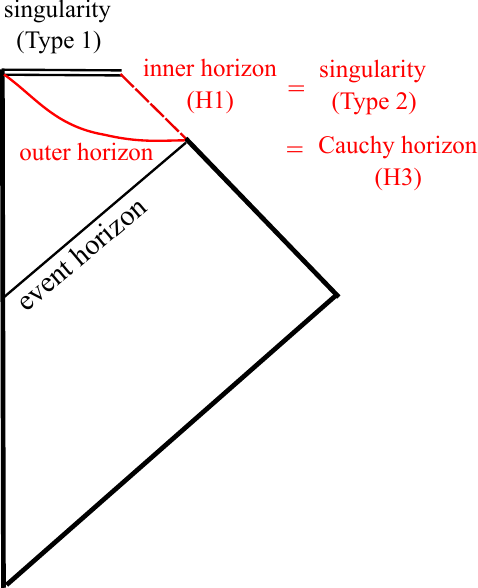}
\caption{\label{fig:charged_MI}Left: Dynamical formation of a charged black hole, where $P = 0$ \cite{Hong:2008mw}. There exists a Type 1 singularity $r = 0$ (thick black curve) and an outer apparent horizon $r_{,v} = 0$ (red curve). As $v$ goes to infinity, there appears an H1 horizon ($r_{,v} = 0$) which is parallel to the ingoing null direction; this will approach a Type 2 curvature singularity ($r > 0$). Here, the grid spacing is $1$ for black contours and $0.1$ for green contours. Right: The Penrose diagram for the dynamical formation of a charged black hole.}
\includegraphics[scale=0.35]{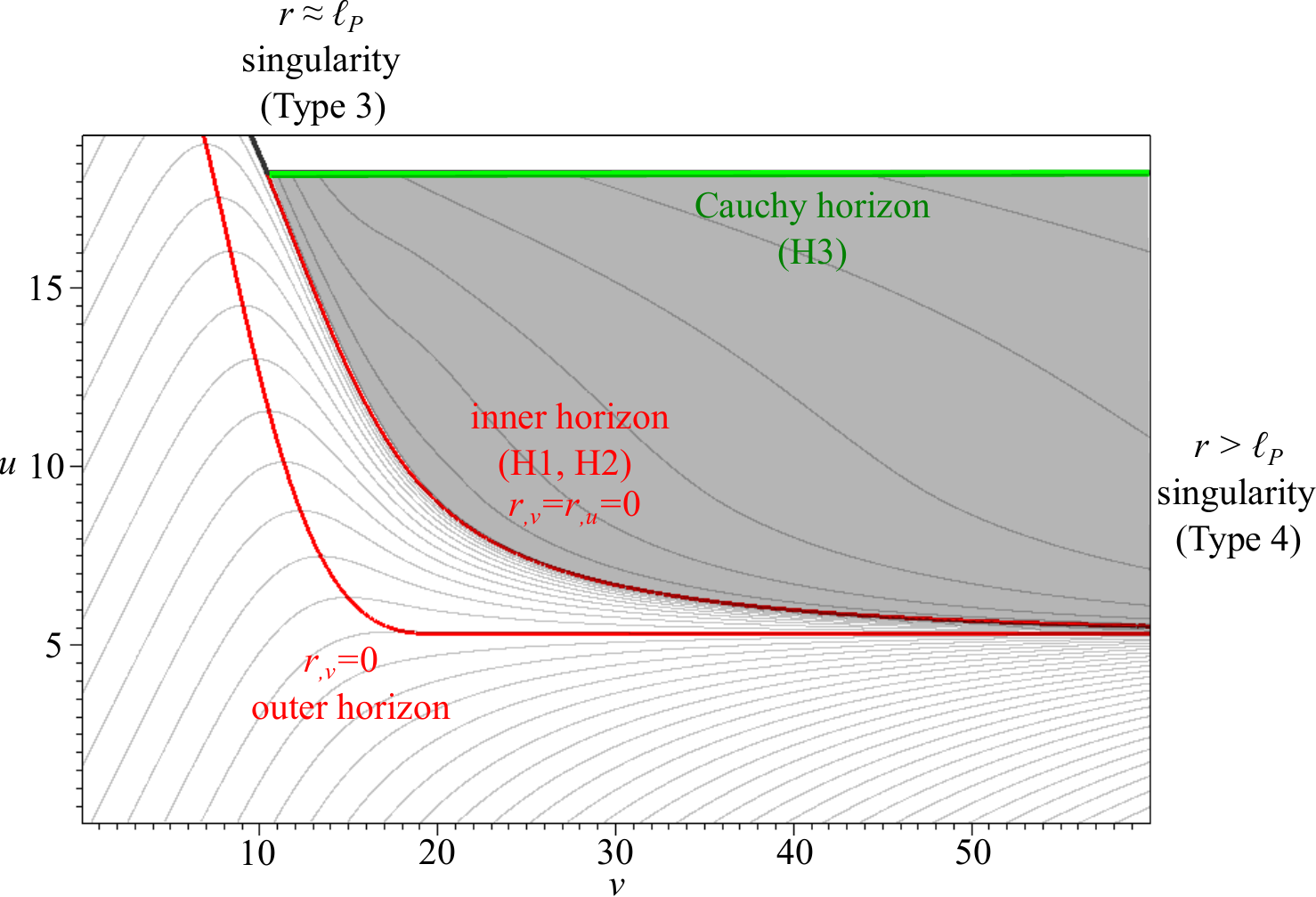}
\includegraphics[scale=0.55]{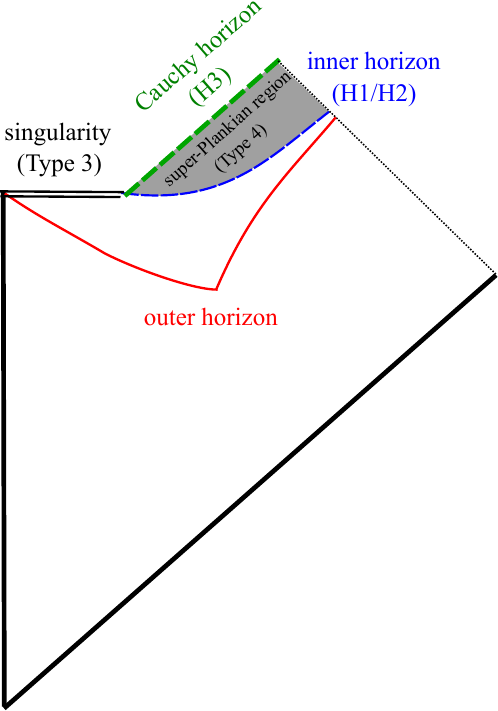}
\caption{\label{fig:charged_HR_G}Left: The evaporating charged black hole, where $P = 0.1$ \cite{Hong:2008mw}. There exists a Type 3 singularity with $r \simeq \ell_{P}$ and an outer apparent horizon at $r_{,v} = 0$. At the inner apparent horizon, an H1 horizon ($r_{,v} = 0$) and an H2 horizon ($r_{,u} = 0$) are overlapped. Between the Type 3 singularity and H1/H2 horizon, there appears an H3 Cauchy horizon; we cannot decide beyond the null surface. Beyond the H1/H2 inner apparent horizon, the curvature rapidly increases, and hence, the gray-colored region is the super-Plankian region which goes beyond the scope of the semi-classical description. Therefore, the gray-colored region corresponds to Type 4 singularity ($r > \ell_{P}$). Right: The Penrose diagram for an evaporating charged black hole.}
\end{center}
\end{figure}

\subsection{Formation of the inner horizon in Einstein-Maxwell theory}

In this subsection, we discuss the formation and evaporation process of charged black holes based on Ref.~\cite{Hong:2008mw}. Hence, we begin with the typical model where a complex scalar field is coupled with a $U(1)$ gauge field that is given as follows:
\begin{eqnarray} \label{pulse_U1}
S = \int \sqrt{-g} dx^{4} \left[ \frac{\mathcal{R}}{16\pi} - \left( \phi_{;\mu} + ie A_{\mu} \phi \right) \left( \bar{\phi}^{;\mu} - ie A^{\mu} \bar{\phi} \right) - \frac{1}{8\pi} F_{\mu\nu}F^{\mu\nu} \right],
\end{eqnarray}
where $\mathcal{R}$ is the Ricci scalar, $\phi$ is a complex scalar field, $A_{\mu}$ is the $U(1)$ gauge field, $e$ is the gauge coupling, and $F_{\mu\nu} = A_{\nu;\mu} - A_{\mu;\nu}$. In the double-null formalism, one can observe a charged black hole that is formed in this model due to the appearance of the input pulse $\phi(u,v)$ at $u=0$ to cause a gravitational collapse to occur in the spacetime,
\begin{eqnarray}
\phi(0,v) = \frac{A}{\sqrt{4\pi}} \sin^{2} \left( \frac{v}{v_{\mathrm{f}}} \right) \exp \left( 2\pi i \frac{v}{v_{\mathrm{f}}} \right),
\end{eqnarray}
where the pulse is defined in $0 \leq v \leq v_{\mathrm{f}}$, $A = 0.25$, and $e = 0.1$.

To turn on the Hawking radiation, we introduce the renormalized energy-momentum tensor $\langle \hat{T}_{\mu\nu} \rangle$ in an approximate form \cite{Davies:1976ei}, where the tensor is proportional to $P \propto N\ell_{P}^{2}$ (the quantity $P$ is precisely defined in Appendix A), where $N$ is the number of scalar field that contribute to the Hawking radiation; this is the natural cutoff if $N$ fields contribute to Hawking radiation \cite{Dvali:2012uq}. If $P = 0$, we turn off Hawking radiation; we choose $P = 0.1$ for the evaporating case.

Solving the equations of motion derived from Eq.~\eqref{pulse_U1} numerically using the double-null formalism (see Appendices A and C for more details), our numerical results demonstrate that a charged black hole is formed and evaporated during the process of gravitational collapse. Regarding this, we address several important remarks from the numerical results in \cite{Hong:2008mw}:
\begin{itemize}
\item[--] 1. There exist two kinds of singularities, which are Type 1 and Type 2. On the left of Fig.~\ref{fig:charged_MI}, one can see a Type 1 singularity at $r = 0$ (thick black curve) and an outer apparent horizon $r_{,v} = 0$ (red curve).

In the late time limit, i.e., as $v$ goes to infinity, there appears the H1 horizon ($r_{,v} = 0$) which is parallel to the ingoing null direction. This will eventually approach a Type 2 curvature singularity ($r > 0$) due to the mass inflation effect.

Therefore, the Type 2 singularity coincides with the horizon H1, although it is parallel with constant $v$ lines. This is somehow counter-intuitive with the \textit{common sense} which has been mentioned in Sec.~\ref{sec:pre}. However, this will help to preserve the strong cosmic censorship conjecture, because no observer can penetrate the Cauchy horizon due to the weak singularity.
\item[--] 2. If we turn on the Hawking radiation as well as the incoming negative energy flux, the situation becomes more complicated. In the left of Fig.~\ref{fig:charged_HR_G}, there exists the Type 3 singularity $r \simeq \ell_{P}$ and an outer apparent horizon $r_{,v} = 0$. At the inner apparent horizon, the H1 horizon ($r_{,v} = 0$) and the H2 horizon ($r_{,u} = 0$) are overlapped. Hence, as an observer penetrates inside the inner apparent horizon, the observer will experience that the areal radius increases, which is analogous to a wormhole structure (see more details in \cite{Hwang:2010im}). Due to the existence of the Type 3 singularity, there appears the H3 Cauchy horizon which is the future affected region from the singularity. Beyond the H1/H2 inner apparent horizon, the curvature rapidly increases, approximately $m \sim e^{+M^{2}}$, where $m$ is the Misner-Sharp mass \cite{Hong:2008mw}; hence, the gray-colored region represents the super-Plankian region which goes beyond the scope of the semi-classical description. Therefore, the gray-colored region (inside the H1 and H2 inner horizons) corresponds to Type 4 singularity ($r > \ell_{P}$).

To summarize, there appears both Type 3 singularity and Type 4 singularity. Type 3 singularity generates the Cauchy horizon, while it is hidden by the Type 4 singularity. On the other hand, if one can trust the computation beyond the Type 4 singularity, e.g., assuming a large number of matter fields ($\sim e^{M^{2}}$, see \cite{Yeom:2009zp,Kim:2013fv}), the strong cosmic censorship might be violated even with the semi-classical effects \cite{Poisson:1997my}. The Type 4 singularity very well coincides with an inner apparent horizon, although there is an ambiguity in defining the exact location of the singularity because it depends on the choice of the cutoff.
\end{itemize}

Therefore, if we only define the apparent horizon using H1, one may neglect the important and interesting behaviors relating to the inner horizon and the Cauchy horizon. We can conclude that in dynamical situations, H1, H2, and H3 horizons are distinguishable; also, Type 1, Type 2, Type 3, and Type 4 singularities all appear in different situations.

As we have mentioned, Fig.~\ref{fig:charged_MI} confirms that the interior of a black hole does not remain as the interior of the static solution. The inner horizon becomes the $r_{,v}= 0$ horizon and is parallel to the in-going null direction, which is inconsistent with the static solution. Moreover, in the late time ($v \rightarrow \infty$) limit, the inner horizon becomes a curvature singularity, and hence, the regularity condition is also not satisfied.

\subsection{Do hairs remove the Cauchy horizon?: A case of the Brans-Dicke theory}

Some literature heavily relied on the assumption that there may exist hairs associated with the matter fields to establish some theorems about the non-existence of Cauchy horizons. In this subsection, we focus on the cases in which some hairs may help to remove the Cauchy horizon, but surprisingly there \textit{exists} a counter-example that there is the Cauchy horizon with a scalar hair. Here, a crucial factor to support the counter-example is the existence of scalar hair at the outer horizon, although the non-trivial scalar hair can exist on the inner horizon.

To investigate this issue, we consider the following model with a non-minimally coupled scalar field \cite{Hansen:2014rua,Hansen:2015dxa,Nakonieczna:2015umf,Nakonieczna:2016iof}:
\begin{equation} \label{BD_action}
S = \int \sqrt{-g} dx^{4} \left[ \frac{1}{16\pi} \left( \Phi \mathcal{R} - \frac{\omega}{\Phi} \Phi_{;\mu} \Phi^{;\mu} - V(\Phi) \right) + \Phi^{\beta} \left( - \frac{1}{2} \left( \phi_{;\mu} + ie A_{\mu} \phi \right) \left( \bar{\phi}^{;\mu} - ie A^{\mu} \bar{\phi} \right) - \frac{1}{16\pi} F_{\mu\nu}F^{\mu\nu}\right) \right],
\end{equation}
where $\Phi$ is the Brans-Dicke field, $\omega$ is the Brans-Dicke parameter, $V(\Phi)$ is the potential of the Brans-Dicke field, and $\beta$ is a constant. Similarly, to observe the formation of the inner horizon for charged black hole during the gravitational collapse using the double-null formalism we send an input pulse of a charged scalar field ($0 \leq v \leq v_{\mathrm{f}}$) at $u=0$, where it is given by
\begin{eqnarray}
\phi(0,v) = \frac{A}{\sqrt{4\pi}} \sin^{4} \left( \frac{v}{v_{\mathrm{f}}} \right) \exp \left( 2\pi i \frac{v}{v_{\mathrm{f}}} \right).
\end{eqnarray}
For simplicity, we fix $e = 0.3$ and $\beta = 0$ or $\beta = 1$.

\begin{figure}
\begin{center}
\includegraphics[scale=0.32]{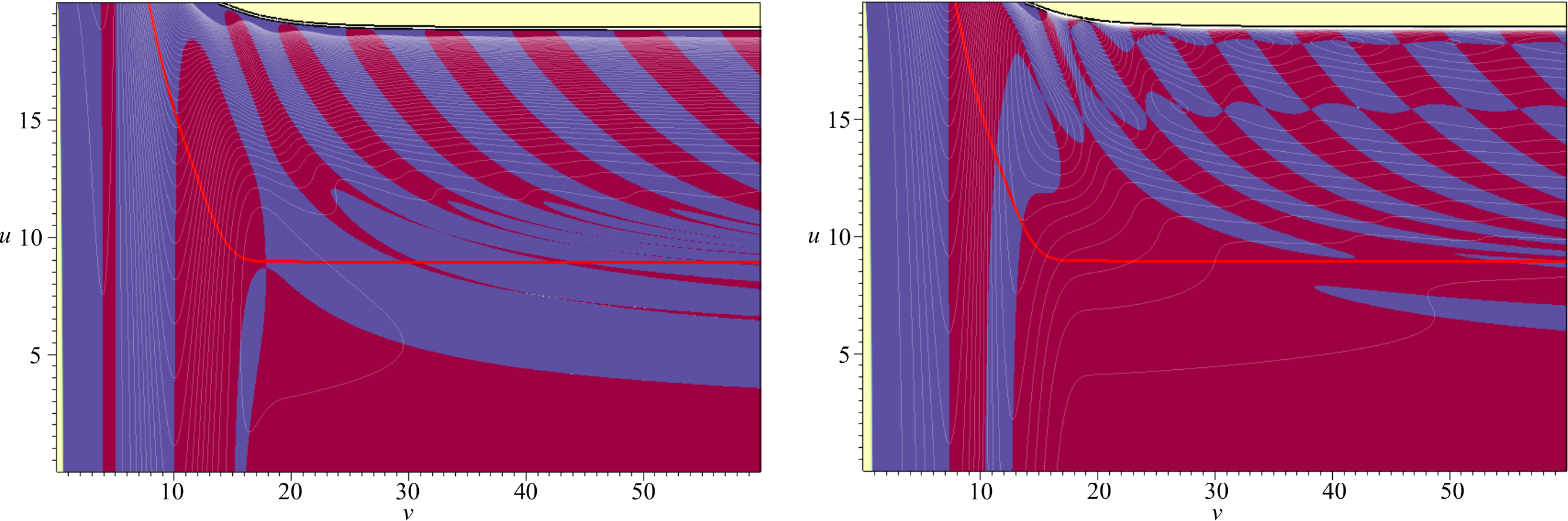}
\caption{\label{fig:charged_hairs}The real part (left) and the imaginary part (right) of the complex scalar field when $\omega = 1000$ and $\beta=0$. Hence, the dynamics of the Brans-Dicke field can be negligible and this corresponds to the Einstein gravity \cite{Nakonieczna:2015umf,Nakonieczna:2016iof}. Here, we used $A = 0.25/\sqrt{2}$. The thin white contours are the contours of the fields and the red curves are apparent horizons ($r_{,v}= 0$). The blue-colored regions are where the contours are space-like; the red-colored regions are where the contours are time-like. This directly shows that the scalar field has non-trivial dynamics in the Cauchy horizon limit.}
\includegraphics[scale=0.19]{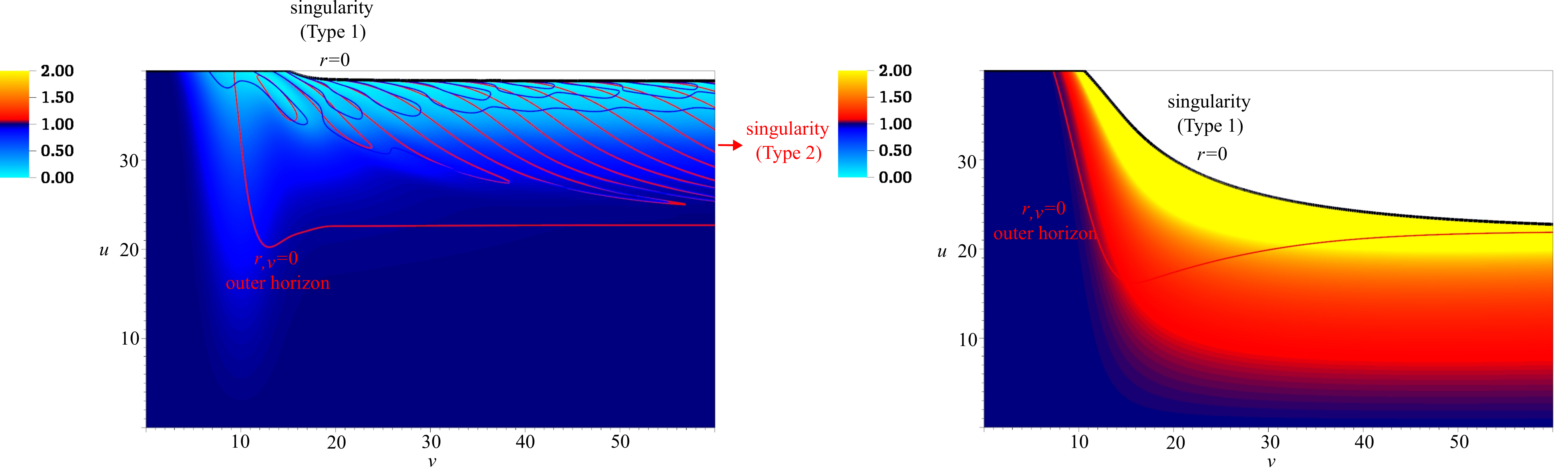}
\caption{\label{fig:hairs}The Brans-Dicke field $\Phi$ and horizons of charged black holes with $\Phi$ when $A=0.15$ and $\omega = -1.4$ \cite{Hansen:2014rua,Hansen:2015dxa}. Left: $\beta = 0$ and hence there is no Brans-Dicke hair at the outer horizon. Inside the outer apparent horizon, $r_{,v}=0$ (red curve) and $r_{,u} = 0$ (light blue curve) horizons appear in a highly sophisticated manner because of the Jordan frame. In the $v \rightarrow \infty$ limit, there exists a Type 2 singularity, but there exists a gradient of the Brans-Dicke field along the $u$ direction. Right: $\beta = 1$ and there exists Brans-Dicke hair along the outer apparent horizon. The corresponding hair causes neither the inner apparent horizon nor Type 2 singularity to exist inside the outer horizon.}
\end{center}
\end{figure}

In this model, we can consider the Brans-Dicke field which is non-minimally coupled with the Maxwell field. Solving the equations of motion derived from Eq.~\eqref{BD_action} numerically in the double-null formalism (see Appendices B and C for more details), we demonstrate our numerical results from \cite{Hansen:2014rua,Hansen:2015dxa,Nakonieczna:2015umf,Nakonieczna:2016iof} in the following.
\begin{itemize}
\item[--] 1. In Fig.~\ref{fig:charged_hairs}, we considered $\beta = 0$ and $\omega = 1000$. The large value of $\omega$ implies that the dynamics of the Brans-Dicke field can be switched off effectively and this result is consistent with Einstein gravity.

Interestingly, the dynamics of the real and imaginary parts of the complex scalar field do not disappear in the large $v$ limit. We find that the derivative of a complex scalar field with respect to $u$ might vanish but might not vanish with respect to $v$. Therefore, the dynamics of a non-trivial complex scalar field can exist in the inner horizon.

If we turn on the Hawking radiation, one can still see the dynamics of the scalar fields that penetrate beyond the inner apparent horizon: see Fig.~15 of \cite{Hong:2008mw}.
\item[--] 2. We considered $\beta = 0$ in the left of Fig.~\ref{fig:hairs}. The minimal coupling between the Maxwell field and the Brans-Dicke field in the Einstein frame does not form Brans-Dicke hair on the outer horizon.

Inside the outer apparent horizon, the appearance of $r_{,v}=0$ (red curve) and $r_{,u} = 0$ (light blue curve) horizons are highly sophisticated because we perform the calculations in the Jordan frame. Hence, both H1 and H2 can appear in various places in the Jordan frame.

In the $v \rightarrow \infty$ limit, there exists a Type 2 singularity, but there exists a gradient of the Brans-Dicke field along the $u$ direction. Therefore, one can interpret that there exists the Cauchy horizon in the $v\rightarrow \infty$ limit, and at the same time, there exists a non-trivial scalar field dynamics along the Cauchy horizon.
\item[--] 3. In the right of Fig.~\ref{fig:hairs}, we considered the $\beta = 1$ case. The non-minimal coupling between the Brans-Dicke field and the Maxwell field gives rise to the existence of the Brans-Dicke hair along the outer apparent horizon.

One interesting observation is that there is no inner apparent horizon or Type 2 singularity inside the outer horizon. So, this implies that \textit{if there exists a hair at the outer horizon, the inner Cauchy horizon structure disappears}, which is very different from pure Einstein gravity. Of course, at this moment, this is just a phenomenological observation and we need further mathematical investigations.
\end{itemize}

Therefore, the scalar hair at the outer apparent horizon plays a more important role than the gradient of the (scalar or vector) field along the inner apparent horizon to determine the existence of the Cauchy horizon.

\section{\label{sec:con}Conclusion}

In this paper, we investigated the properties of inner horizons. First, we distinguished two notions of horizons, where one is the quasi-local notion (e.g., apparent horizons) and another is the global notion (Cauchy horizon). For the quasi-local notion, we can distinguish two horizons, i.e., $r_{,v} = 0$ or $r_{,u} = 0$. For the global notion, we can distinguish the horizons whether the singularity was generated from the classical singularity or the quantum (Planckian) singularity. In fully dynamic cases, one can construct several models so that all the notions can be distinguished. 

Our results provide a few important implications for the cosmic censorship conjecture. How can we define the singularity inside a black hole, for example, in the classical or quantum sense? What are the conditions to rescue the cosmic censorship conjecture, for example, in the classical or quantum sense? If there is no Cauchy horizon, is it related to the quasi-local or the global notion of the horizons? To preserve the cosmic censorship conjecture, should the horizon be disappeared at all or the existence of the inner horizon itself is still fine? These are the diverse questions that we need to ask before we proceed with more proof of the non-existence of the Cauchy horizon. We will leave these fascinating topics for future investigation and hope that our paper can shed light on the future advances for this issue.

In addition, we can provide some guidelines for the Cauchy horizon theorems. Our simulations demonstrate that when we consider black holes in the dynamical cases, their interior structure is no longer to be described by the static and regular metric. Also, the scalar hair on the outer horizon is more important than the scalar hair on the inner horizon to determine the existence of the Cauchy horizon. We need to investigate these issues using not only numerical but also analytic approaches. Will this behavior generally be true in most cases? We leave this interesting question to be investigated in the future.

\newpage

\section*{Acknowledgment}

We would like to thank Mu-In Park for his valuable comments. DY is supported by the National Research Foundation of Korea (Grant no.: 2021R1C1C1008622, 2021R1A4A5031460). XYC acknowledges the support from the starting grant of Jiangsu University of Science and Technology (JUST).

\section*{Appendix A: Charged black holes in Einstein gravity}

We start from the model with a complex massless scalar field $\phi$ and a gauge field $A_{\mu}$ \cite{Hong:2008mw}:
\begin{eqnarray}
S = \int \sqrt{-g} dx^{4} \left[ \frac{\mathcal{R}}{16\pi} - \left( \phi_{;\mu} + ie A_{\mu} \phi \right) \left( \bar{\phi}^{;\mu} - ie A^{\mu} \bar{\phi} \right) - \frac{1}{8\pi} F_{\mu\nu}F^{\mu\nu} \right],
\end{eqnarray}
where $F_{ab}=A_{b;a}-A_{a;b}$ and $e$ is the gauge coupling. We introduce the double-null coordinates
\begin{eqnarray}
ds^{2} = - \alpha^{2}(u,v) du dv + r^{2}(u, v) d\Omega_{2}^{2},
\end{eqnarray}
and define variables: 
\begin{eqnarray}
h \equiv \frac{\alpha_{,u}}{\alpha},\quad d \equiv \frac{\alpha_{,v}}{\alpha},\quad f \equiv r_{,u},\quad g \equiv r_{,v},\quad w \equiv \sqrt{4\pi} \phi_{,u},\quad z \equiv \sqrt{4\pi} \phi_{,v}.
\end{eqnarray}
In addition, we assume the Coulomb gauge $A_{\mu} = (a(u,v), 0, 0, 0)$ which allows us to define the charge function $q(u,v) \equiv 2r^{2} a_{,v} / \alpha^{2}$.

The semi-classical Einstein equation is defined as
\begin{eqnarray}
G_{\mu\nu}=8\pi \left( T^{\mathrm{C}}_{\mu\nu}+\langle \hat{T}^{\mathrm{H}}_{\mu\nu} \rangle \right),
\end{eqnarray}
where $T^{\mathrm{C}}_{\mu\nu}$ is the classical energy-momentum tensor and $\langle \hat{T}^{\mathrm{H}}_{\mu\nu} \rangle$ is the renormalized energy-momentum tensor. We introduce the $S$-wave approximation of the renormalized energy-momentum tensor:
\begin{eqnarray}
\langle \hat{T}^{\mathrm{H}}_{uu} \rangle &=& \frac{P}{4\pi r^{2}}\left(h_{,u}-h^{2}\right),
 \\
\langle \hat{T}^{\mathrm{H}}_{uv} \rangle = \langle \hat{T}^{\mathrm{H}}_{vu} \rangle &=& -\frac{P}{4\pi r^{2}}d_{,u},
 \\
\langle \hat{T}^{\mathrm{H}}_{vv} \rangle &=& \frac{P}{4\pi r^{2}}\left(d_{,v}-d^{2}\right),
\end{eqnarray}
with $P \equiv N \ell_{\mathrm{P}}^2 / 12\pi$, where $N$ is the number of scalar field and $\ell_{\mathrm{P}}$ is the Planck length.

The semi-classical Einstein equations, as well as the scalar field equation ($s \equiv \sqrt{4\pi}\phi$) and the Maxwell equation, are summarized as follows:
\begin{eqnarray}
d_{,u} = h_{,v} &=& \frac{1}{1-P/r^{2}} \left[ \frac{fg}{r^{2}} + \frac{\alpha^2}{4r^{2}} - \frac{\alpha^{2} q^{2}}{2 r^{4}} - \frac{1}{2}(w\overline{z}+\overline{w}z) - \frac{iea}{2}(s\overline{z}-\overline{s}z) \right], \\
g_{,v} &=& 2dg - rz\overline{z} - \frac{P}{r} \left( d_{,v}-d^{2} \right),\\
f_{,u} &=& 2fh - rw\overline{w} - iear\left(\overline{w}s-w\overline{s}\right) - e^{2}a^{2}rs\overline{s} - \frac{P}{r} \left(h_{,u}-h^{2}\right),\\
f_{,v} = g_{,u} &=& -\frac{fg}{r} - \frac{\alpha^{2}}{4r} + \frac{\alpha^{2} q^{2}}{4 r^{3}} - \frac{P}{r} d_{,u}, \\
a_{,v} &=& \frac{\alpha ^{2} q}{2 r^{2}}, \\
q_{,v} &=& -\frac{ier^{2}}{2} \left( \overline{s}z-s\overline{z} \right), \\
z_{,u} = w_{,v} &=& - \frac{fz}{r} - \frac{gw}{r} - \frac{iearz}{r} - \frac{ieags}{r} - \frac{ie}{4r^{2}}\alpha^{2}qs.
\end{eqnarray}

\section*{Appendix B: Charged black holes in Brans-Dicke gravity}

We consider the model for the charged black holes in Brans-Dicke gravity \cite{Hansen:2014rua,Hansen:2015dxa}
\begin{equation}
S = \int \sqrt{-g} dx^{4} \left[ \frac{1}{16\pi} \left( \Phi \mathcal{R} - \frac{\omega}{\Phi} \Phi_{;\mu} \Phi^{;\mu} - V(\Phi) \right) + \Phi^{\beta} \left( - \frac{1}{2} \left( \phi_{;\mu} + ie A_{\mu} \phi \right) \left( \bar{\phi}^{;\mu} - ie A^{\mu} \bar{\phi} \right) - \frac{1}{16\pi} F_{\mu\nu}F^{\mu\nu}\right) \right],
\end{equation}
where $\Phi$ is the Brans-Dicke field, $\omega$ is the Brans-Dicke parameter, $V(\Phi)$ is the potential of the Brans-Dicke field, and $\beta$ is a constant. We define the additional variables:
\begin{eqnarray}
W \equiv \Phi_{,u},\quad Z \equiv \Phi_{,v}.
\end{eqnarray}

The Einstein equations, as well as the Brans-Dicke field equation, the complex scalar field equation, and the Maxwell equation, are given by the following:
\begin{eqnarray}
d_{,u} = h_{,v} &=& \mathfrak{A} - \frac{\mathfrak{B}}{r} - \frac{\mathfrak{C}}{2 r \Phi}, \\
f_{,v} = g_{,u} &=& \mathfrak{B} - \frac{\mathfrak{C}}{2\Phi},\\
Z_{,u} = W_{,v} &=& \frac{\mathfrak{C}}{r},\\
\label{eq:E1}r_{,uu} &=& 2fh - \frac{r}{2 \Phi} \left( W_{,u}-2hW \right) - \frac{r \omega}{2 \Phi^{2}} W^{2} - 4 \pi r \Phi^{\beta - 1} {T}^{\mathrm{C}}_{uu},\\
\label{eq:E2}r_{,vv} &=& 2gd - \frac{r}{2 \Phi} \left( Z_{,v}-2dZ \right) - \frac{r \omega}{2 \Phi^{2}} Z^{2} - 4 \pi r \Phi^{\beta -1} {T}^{\mathrm{C}}_{vv},\\
\label{eq:a1}a_{,v} &=& \frac{\alpha ^{2} q}{2 r^{2}}, \\
\label{eq:q1}q_{,v} &=& -\frac{ier^{2}}{2} \left( \overline{s}z-s\overline{z} \right) - \beta q \frac{Z}{\Phi}, \\
\label{eq:avv}a_{,vv} &=& \frac{\alpha^{2}}{r^{2}} \left( d - \frac{g}{r} \right)q - \frac{ie\alpha^{2}}{4} \left( z\overline{s}-s\overline{z}\right) - \beta q \frac{\alpha^{2} Z}{2r^{2}\Phi},\\
q_{,u} &=& \frac{ier^{2}}{2} \left( \overline{s}w-s\overline{w} \right) - r^{2}e^{2}a s \overline{s} - \beta q \frac{W}{\Phi}, \\
\label{eq:a2}a_{,uv} &=& \frac{\alpha^{2}}{r^{2}} \left( h - \frac{f}{r} \right)q + \frac{ie\alpha^{2}}{4} \left( w\overline{s}-s\overline{w}\right) - \frac{\alpha^{2}}{2}e^{2}as\overline{s} - \beta q \frac{\alpha^{2} W}{2r^{2}\Phi},\\
\label{eq:s}s_{,uv} &=& - \frac{fz}{r} - \frac{gw}{r} - \frac{iearz}{r} - \frac{ieags}{r} - \frac{ie}{4r^{2}}\alpha^{2}qs - \frac{\beta}{2\Phi} \left( Wz +Zw + ies a Z \right),
\end{eqnarray} 
where
\begin{eqnarray}
\label{eq:A}\mathfrak{A} &\equiv& -\frac{2\pi \alpha^{2}}{r^{2}}\Phi^{\beta - 1} T^{\mathrm{C}}_{\theta\theta} - \frac{1}{2r \Phi} \left( gW+fZ \right) -\frac{\omega}{2\Phi^{2}}WZ, \\
\label{eq:B}\mathfrak{B} &\equiv& - \frac{\alpha^{2}}{4r} - \frac{fg}{r} + 4 \pi r \Phi^{\beta-1} T^{\mathrm{C}}_{uv} - \frac{1}{\Phi}(gW+fZ), \\
\label{eq:C}\mathfrak{C} &\equiv& - fZ - gW - \frac{2\pi r \alpha^{2} \Phi^{\beta}}{3+2\omega} \left( T^{\mathrm{C}} - 2 \beta {\mathcal{L}}^{\mathrm{EM}} \right),\\
\label{eq:TSuu}T^{\mathrm{C}}_{uu} &\equiv& \frac{1}{4\pi} \left[ w\overline{w} + iea(\overline{w}s-w\overline{s}) +e^{2}a^{2}s\overline{s} \right],\\
\label{eq:TSuv}T^{\mathrm{C}}_{uv} &\equiv& \frac{{(a_{,v})}^{2}}{4\pi\alpha^{2}},\\
\label{eq:TSvv}T^{\mathrm{C}}_{vv} &\equiv& \frac{1}{4\pi} z\overline{z},\\
\label{eq:TSthth}T^{\mathrm{C}}_{\theta\theta} &\equiv& \frac{r^{2}}{4\pi\alpha^{2}} \left[ (w\overline{z}+z\overline{w}) + iea(\overline{z}s-z\overline{s})+\frac{2{(a_{,v})}^{2}}{\alpha^{2}} \right],\\
T^{\mathrm{C}} &\equiv& - \frac{4}{\alpha^{2}}T^{\mathrm{C}}_{uv} + \frac{2}{r^{2}}T^{\mathrm{C}}_{\theta \theta},\\
\mathcal{L}^{\mathrm{EM}} &\equiv& \frac{1}{4\pi\alpha^{2}} \left( w\bar{z} + z\bar{w} \right) + \frac{iea}{4\pi\alpha^{2}} \left( \bar{z}s - z\bar{s}\right) + \frac{{a_{,v}}^{2}}{2\pi \alpha^{4}}.
\end{eqnarray}

\section*{Appendix C: Boundary conditions, integration algorithm, and numerical convergence test}

Continuing from Appendix B, we need to provide the initial conditions for all functions ($\alpha, r, g, f, h, d, s, w, z, a, q, \Phi, W, Z$) on the initial surfaces $u=0$ and $v=0$ to solve all equations.

Using the gauge freedom to choose the initial $r$ function, we define $r(0,v) = r_{0} + 0.5 v$ and $r(u,0) = r_{0} - 0.5 u$, where $r_{0}$ is an arbitrary constant. We used $r_{0} = 10$ in the Einstein gravity case and $r_{0} = 20$ in the Brans-Dicke gravity case. This provides $g(0,v) = 0.5$ and $f(u,0) = - 0.5$. We have specified the form of the scalar field profile $s(0,v)$ along the $u = 0$ surface, while it vanishes along the $v = 0$ surface. In addition, we fix $\Phi(u,0) = \Phi(0,v) = 1$ in the Brans-Dicke model. This automatically fixes $z(0,v) = \phi_{,v}(0,v)$, $w(u,0) = \phi_{,u}(u,0)$, $Z(0,v) = 0$, and $W(u,0) = 0$. Hence, $q(u,0) = a(u,0) = 0$ and $\alpha(u,0)=1$. Other functions will be determined by the equations with $uu$- or $vv$- derivation equations (Eqs.~\eqref{eq:E1}, \eqref{eq:E2}, \eqref{eq:avv}). Interested readers can refer to \cite{Nakonieczna:2018tih} on the generic procedure to assign the boundary conditions.

\begin{figure}
\begin{center}
\includegraphics[scale=1]{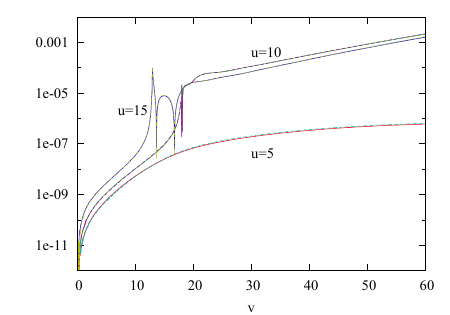}
\caption{\label{fig:convergence}An example of the convergence test of Fig.~\ref{fig:charged_HR_G} \cite{Hong:2008mw}. Here, plots of errors with different step sizes are denoted. We compared $|r_{(1)}- r_{(2)}|/r_{(2)}$ (solid curves) and $4|r_{(2)} - r_{(4)}|/r_{(4)}$ (dashed curves) along a few constant $u$ lines, where $r_{(n)}$ is calculated in an $n$-times finer simulation than $r_{(1)}$.}
\end{center}
\end{figure}

Regarding the convergence test of numerical calculations, interested readers can also refer to the appendices in the following references for detailed description \cite{Hong:2008mw,Hansen:2014rua,Hansen:2015dxa}. Here Fig.~\ref{fig:convergence} shows the error for the case of charged black holes in Brans-Dicke gravity, the error inside the horizon ($u=10, 15$) is still sufficiently small, while the convergence is the second order (this depends on the detailed algorithm).

\end{document}